# HIGH PRESSURE STUDIES OF $T_c$ AND LATTICE PARAMETERS OF $MgB_2$

Sonja I. Schlachter[1], Walter H. Fietz[1], Kai Grube[2,3], Wilfried Goldacker[1]

[1] Forschungszentrum Karlsruhe - Institut für Technische Physik
P.O. Box 3640, 76021 Karlsruhe, Germany.

[2] Forschungszentrum Karlsruhe - Institut für Festkörperphysik,
P.O. Box 3640, 76021 Karlsruhe, Germany.

[3] Current Adress: Cavendish Laboratory, Cambridge CB3 OHE, United Kingdom.

## ABSTRACT

We performed ac-susceptibility measurements of magnesium diboride powder samples under pure hydrostatic pressures up to 0.4 GPa and under quasi-hydrostatic pressure conditions up to 8 GPa in a helium gas pressure cell and a diamond-anvil cell (DAC), respectively. Furthermore, the DAC has been used to investigate the lattice compression by X-ray diffraction. Under helium gas pressure the superconducting transition temperature shows a linear decrease $dT_c/dp$ = -1.13 K/GPa. The pressure effect is fully reversible with a return to the initial $T_c$ of 37.5 K after pressure release. In the diamond-anvil cell the application of pressure leads to a stronger $T_c$-decrease than in the helium gas pressure cell, and after the release of pressure a degradation effect with lower $T_c$ and a broader transition compared to the first measurement at ambient pressure occurs. Such degradation may be explained by shear stresses and uniaxial pressure components, which cannot be avoided in a DAC at low temperatures. Considering the anisotropic $AlB_2$-type structure with alternating honeycomb boron and magnesium layers, $MgB_2$ seems to be very sensitive to such non-hydrostatic pressure components, which would explain the spread of $dT_c/dp$ values reported in the literature.

## INTRODUCTION

The discovery of superconductivity in the diboride $MgB_2$ [1,2] has attracted a lot of attention due to its high transition temperature, which seems to be at or beyond the expected limit of conventional BCS superconductivity with a phonon-mediated pairing mechanism. However, the light masses in $MgB_2$ could lead to high phonon frequencies and therefore, according to BCS theory, to high transition temperatures. The large boron isotope effect on $T_c$, reported by Bud'ko *et al.* [3], points to a phonon-mediated coupling mechanism, too. On the other hand, there are also predictions favoring unconventional superconductivity like the theory of hole superconductivity [4]. In contrast to the expectations of a BCS-like pairing mechanism, this theory predicts positive $T_c$-changes under lattice compression or, strictly speaking, at least positive $T_c$-changes for an in-plane compression of the boron layer.

Due to the anisotropic $AlB_2$-type structure with honeycomb magnesium and boron layers, uniaxial pressure in *a*- or *c*-axis direction could lead to different $T_c$-changes. At the moment, at least to our knowledge, no direct measurements of the uniaxial pressure effects are available due to the small size of the single crystals prepared up to now. However, the knowledge of the axis compressibilities can help to interpret the results of $T_c$ measurements under pressure, which motivates our investigation of both, $T_c$ and lattice parameters of $MgB_2$ presented in this paper.

## EXPERIMENTAL

We have performed ac-susceptibility measurements in a diamond-anvil cell (DAC) and a helium gas pressure cell (HePC) up to pressures of 8 and 0.4 GPa, respectively.

In the DAC the ac-susceptibility of the sample was measured with a pick-up-coil technique that we have especially adapted to our DAC using numerical calculations to find the optimal design of coils and winding numbers. To compensate the unavoidable non-ideal winding of the 20 μm copper wire we included additional compensation coils which are also used to minimize the background signal from the inconel gasket and the steel support of the DAC. Despite the small sample volume and the difficult geometry the large resulting signal to noise ratio allows to monitor the superconducting transition in the real part of the ac-susceptibility, $\chi'_{ac}$, and in the imaginary part, $\chi''_{ac}$, even when powdered material with grain sizes of approximately 10 μm is used.

Commercially available $MgB_2$ powder ('Chempur', Karlsruhe, Germany) was characterized by X-ray diffraction analysis revealing only small amounts of MgO as a secondary phase. Some powder was filled in the sample chamber (0.5 mm diameter, 0.2 mm height) of the DAC together with NaF, serving as a soft and quasi-hydrostatic pressure-transmitting medium. For pressure determination via the ruby-fluorescence method some ruby pieces were added. In parallel the pressure was determined by evaluating the shift of the X-ray diffraction peaks of NaF. Pressure determination was always performed at 37.5 K, i.e. the temperature of the superconducting transition at ambient pressure.

The determination of the lattice parameters under pressure was performed in a second experiment at 300 K in the same experimental setup. In this experiment we used NaCl as pressure transmitting medium and pressure gauge, because, at least at lower pressures, the NaCl reflections do not overlap with any $MgB_2$ reflexes. At pressures above 5 GPa some reflections of the much softer NaCl run through $MgB_2$ reflections making the analysis of the data much more difficult than at lower pressures.

The ac-susceptibility measurements of $MgB_2$ in the HePC were carried out with helium as pressure-transmitting medium. In order to avoid a solidification of the helium at $T \geq 35$ K where the superconducting transition was expected to occur, the pressure was monitored by a manganin gauge and kept below 0.42 GPa throughout the entire measurement.

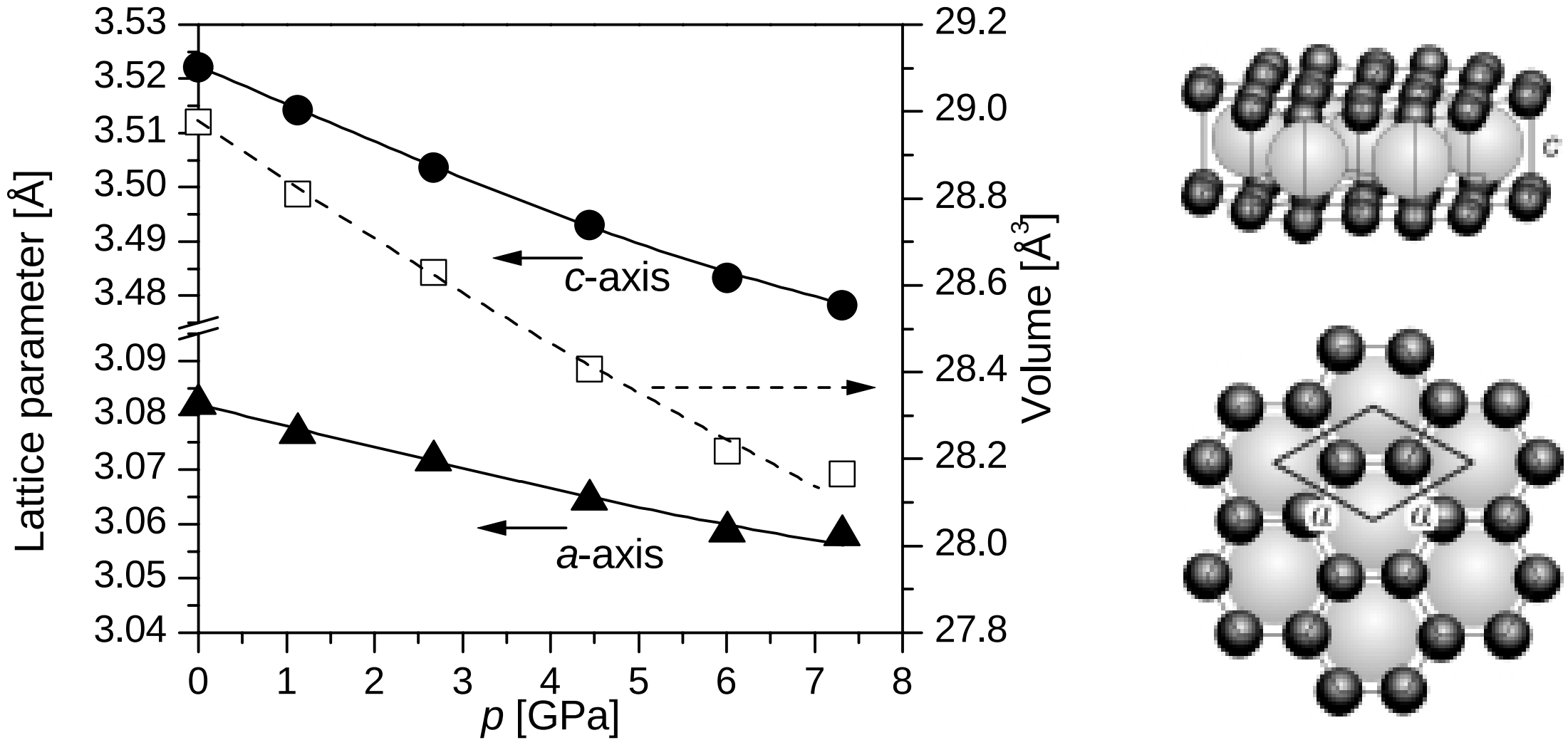

**FIGURE 1:** Lattice parameters and unit-cell volume of $MgB_2$ versus pressure. On the right side schematic pictures of the $AlB_2$-type structure of $MgB_2$ with honeycomb boron layers are presented.

## RESULTS AND DISCUSSION

In the first part of this section we present measurements of the lattice parameters of $MgB_2$ as a function of pressure. The second part is dedicated to the $T_c(p)$ measurements performed in the diamond-anvil cell and in the gas-pressure cell.

### Lattice parameters under pressure

The X-ray diffraction studies on the commercial $MgB_2$ powder were performed in the DAC setup at 300 K using a molybdenum X-ray tube together with a position-sensitive detector. For each pressure six X-ray spectra were taken, each with the DAC turned by a small angle relative to X-ray tube and detector in order to account for the small number of randomly oriented $MgB_2$ grains in the DAC.

FIG. 1 shows the lattice parameters and unit-cell volume of $MgB_2$ as a function of pressure and also the $AlB_2$-type structure with honeycomb boron and magnesium layers. At ambient pressure the lattice parameters are $a_0$ = 3.082 Å and $c_0$ = 3.522 Å, in very good agreement with numerous literature data. Under pressure the *c*-axis decreases stronger than the *a*-axis, indicating that the out-of plane compression is stronger than the in-plane compression. The linear compressibilities are $\kappa_a = 1.6\ 10^{-3}\ \mathrm{GPa}^{-1}$ and $\kappa_c = 2.2\ 10^{-3}\ \mathrm{GPa}^{-1}$. This corresponds with a strong anisotropic thermal expansion being twice as high for the *c*-axis compared to the *a*-axis [5].

From a Birch-fit to the pressure dependence of the unit-cell volume we obtained the bulk modulus $B(p) = B_0 + B' \cdot p$ with $B_0$ = 196 GPa and $B'$=10. This means that $MgB_2$ is a rather hard material. The $B'$ that we obtained for our data is higher than the commonly observed value $B'$ = 4. This unusual high $B'$ parameter that describes the hardening of the material under pressure might be a consequence of the anisotropic structure with softer ionic Mg-B bonds in *c*-axis direction and harder covalent bonds within the boron layer.

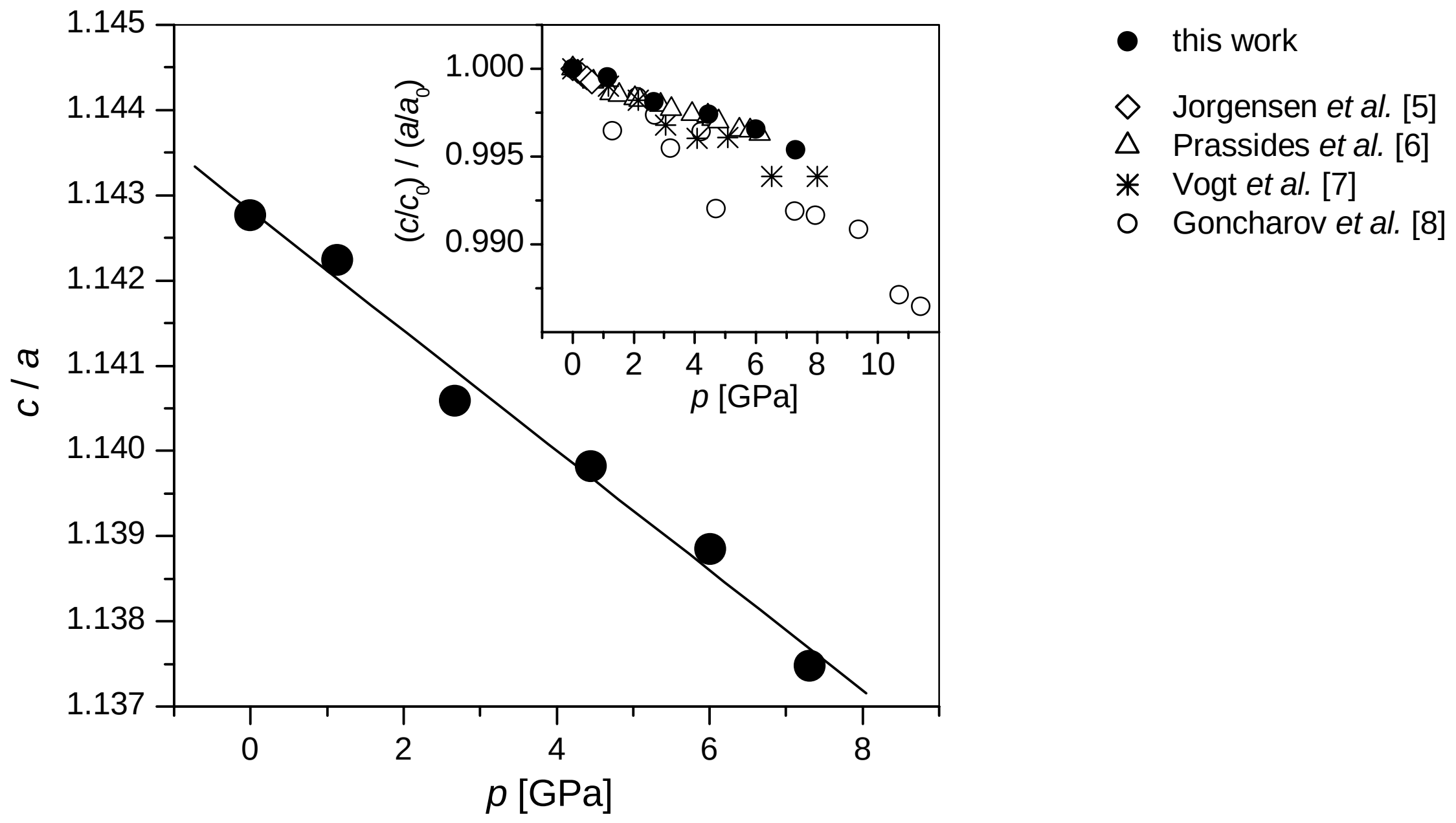

**FIGURE 2:**
Pressure dependence of the *c*/*a*-relation. The inset shows a comparison of the normalized *c*/*a*-relation from our data with data from the literature.

In FIG. 2 we show the pressure dependence of the *c*/*a*-relation. From a linear fit to the data we obtained a decrease of $\mathrm{d}(c/a)/\mathrm{d}p = 7{\cdot}10^{-4}\ \mathrm{GPa}^{-1}$. A comparison of the normalized *c*/*a*-relation from our data to data from the literature is shown in the inset of FIG. 2. The data of Prassides *et al.* [6] are in good agreement with our results whereas the data of Jorgensen *et al.* [5], Vogt *et al.* [7] and Goncharov *et al.*[8] show even larger decreases under pressure.

## $T_c$ as a function of pressure

FIG. 3 presents the results of the $T_c$-measurements performed in the (DAC). FIG. 3a and 3b show the shift of the real part of the ac-susceptibility $\chi'_{ac}$ with increasing and decreasing pressures, respectively. With increasing pressure, the transitions to superconductivity do not only shift to lower temperatures, but also show a strong broadening, and the imaginary part of the ac-susceptibility, which is shown for three different pressures in the inset of FIG. 3a splits in two separated peaks pointing to uniaxial pressure contributions or sample inhomogeneities. The centers of the peaks exhibit different, linear pressure dependencies of –1.3 K/GPa and –1.9 K/GPa. However, when the pressure is released the transitions to superconductivity shift almost parallel to higher temperatures. After the complete release of the pressure the transition to superconductivity is still broadened and occurs at lower temperatures than the initial transition before pressure application. The inset of FIG. 3b shows $T_c$, defined as the temperature where $\chi'_{ac}$ is decreased to 50% of its value in the superconducting state, as a function of pressure. The $T_c$-difference at ambient pressure before and after pressure application and release is approximately 4 K.

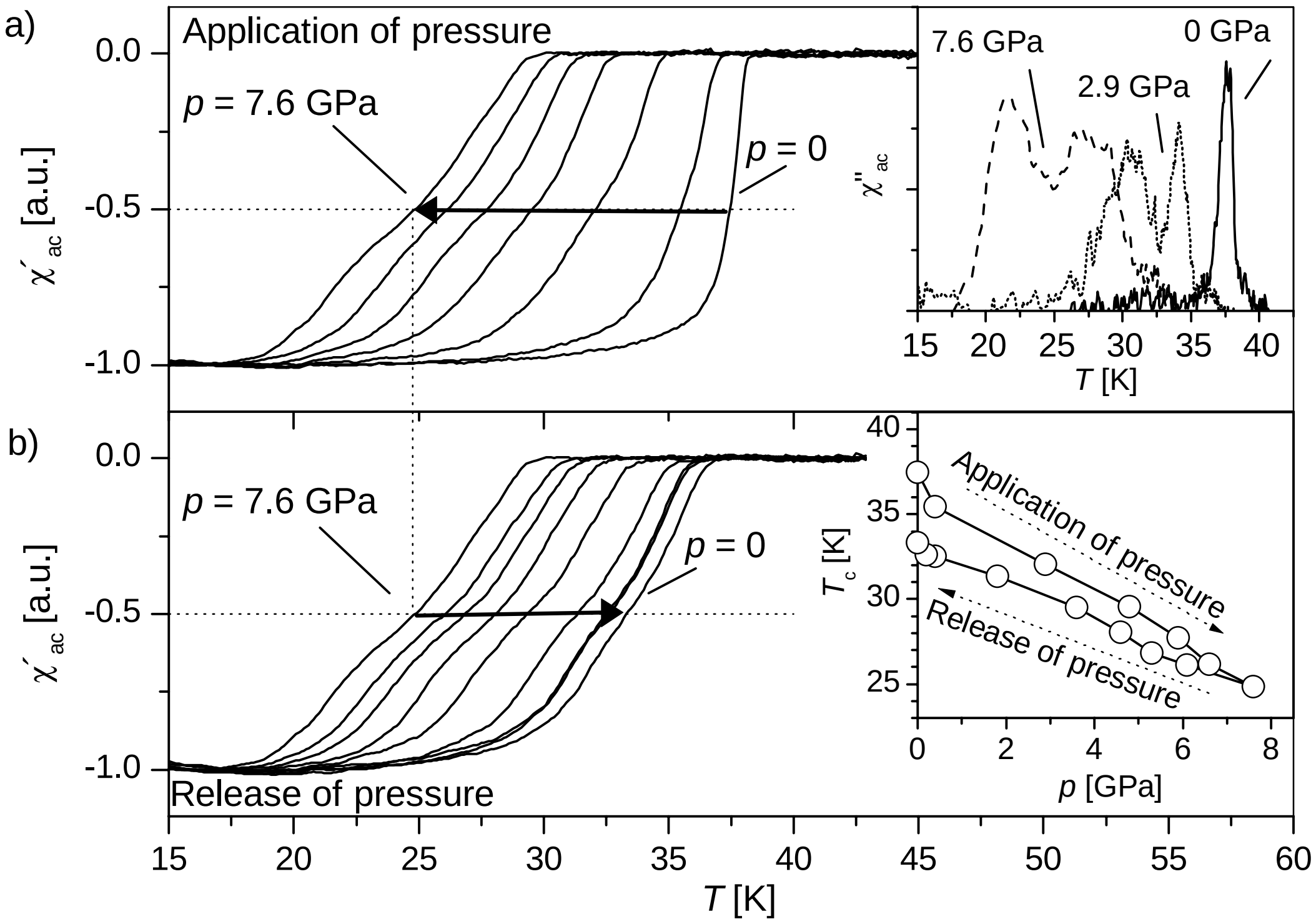

**FIGURE 3**: a) Real part of the ac-susceptibility of $MgB_2$ powder versus temperature measured in the DAC at increasing pressures. Inset: Imaginary parts of the ac-susceptibility at three distinct pressures. b) Shift of the real part of the ac-susceptibility when pressure is released. Inset: $T_c$ versus pressure for pressure application and release.

To ensure that the change of $T_c(p=0)$ during the experiment is not due to experimental problems, we decided to remeasure the few μg powder used in the DAC at ambient pressure in our He gas-pressure cell. FIG. 4 shows the resulting $\chi_{ac}(T)$ from the experiments in the DAC and the HePC using virgin powder and the powder after the release of DAC pressure. From this comparison it is obvious, that the degradation of $T_c(p=0)$ after pressure application in the DAC is rather a result of sample degradation than of experimental problems. We decided to perform $T_c(p)$ measurements in the gas-pressure cell where purely hydrostatic pressure conditions above 35 K can be guaranteed up to pressures of approximately 0.42 GPa (limited by the solidification of He under pressure). The first measurement was performed at ambient pressure giving $T_c = 37.50$ K. Then the maximum pressure $p = 0.42$ GPa was applied at 53 K and the ac-susceptibility measurement was performed from 20 to 53 K. During warming the helium melted at approximately 32.5 K while the transition from the superconducting to the normal state was observed around $T_c = 37.04$ K and $p = 0.41$ GPa. Under pressure the transition in $\chi'_{ac}(T)$ did not show any signs of broadening compared to $\chi'_{ac}(T)$ at ambient pressure. The pressure was released in 5 steps with subsequent measurements of the ac-susceptibility (FIG. 5). After the complete release of the applied pressure $T_c$ returned to 37.49 K, which is, within the experimental error, in very good agreement with the initial value.

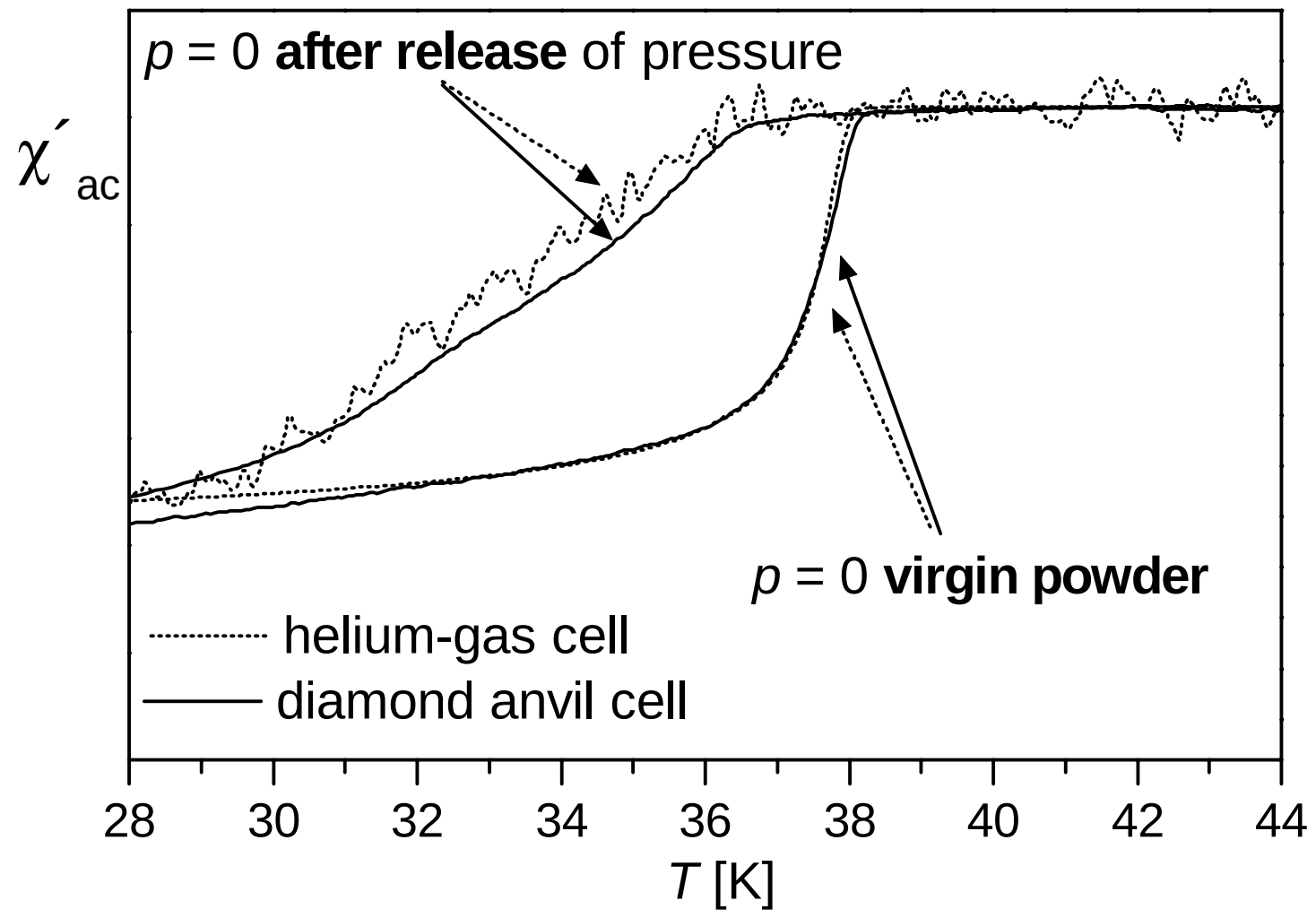

**FIGURE 4**: Real part of the ac-susceptibility of the few μg $MgB_2$ powder before and after the experiment in the DAC (solid line). For comparison we measured the few μg sample before and after the experiment in DAC in our HePC (dotted line). For each state of the sample (virgin and pressure treated) both experiments give within error limits identical results suggesting that the pressure treatment in the DAC causes a clear degradation of the sample with a broader transition and lower $T_c$.

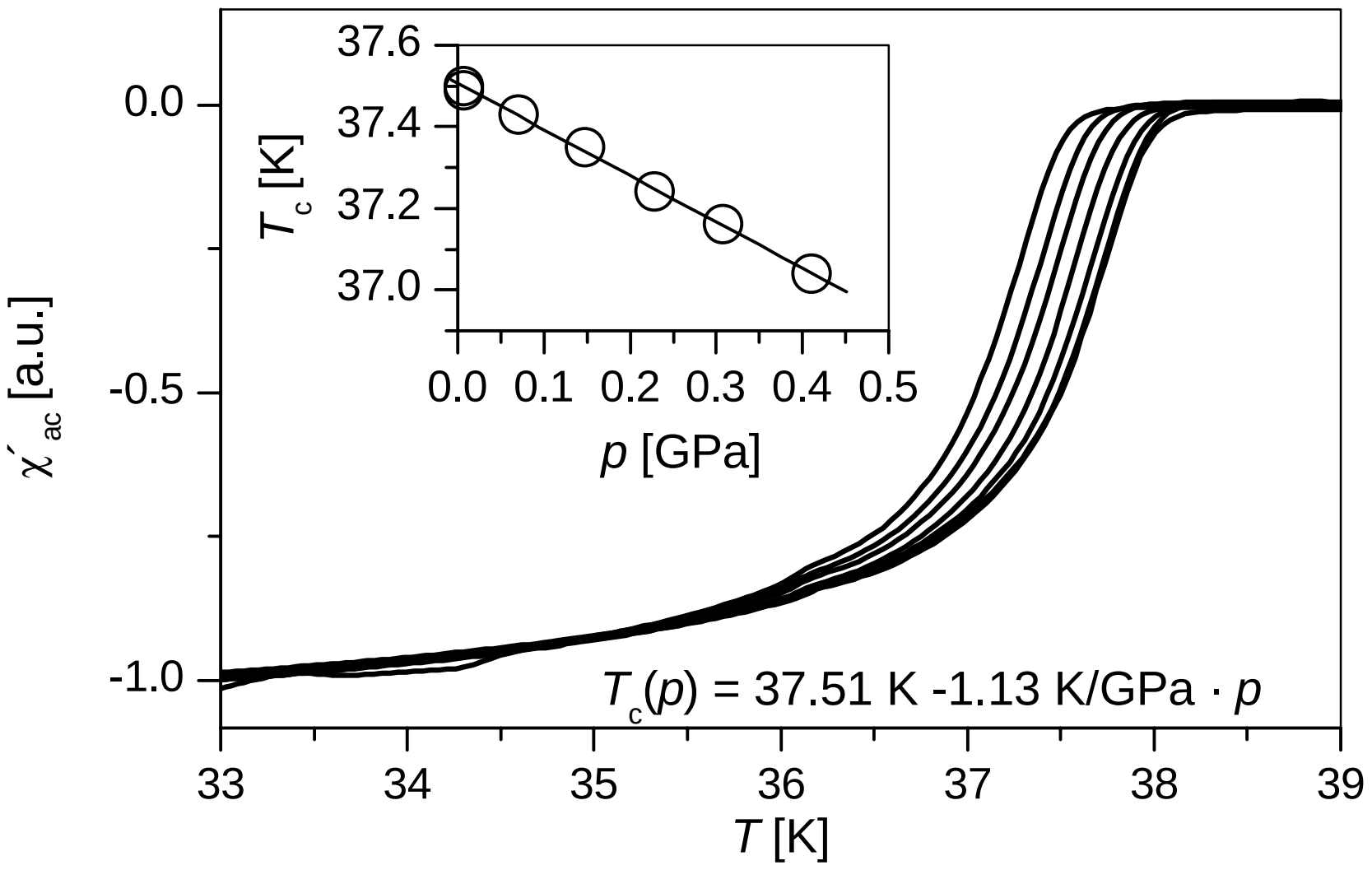

**FIGURE 5:** Real part of the ac-susceptibility versus temperature under purely hydrostatic pressure. The inset shows the decrease of $T_c$ with increasing pressure.

FIGURE 6 shows the $T_c(p)$ data obtained from the experiments with the DAC and the HePC. When pressure is applied the transition temperature in the DAC decreases much faster ($dT_c/dp \approx$ -1.6 K/GPa) than in the gas-pressure cell where we found a pressure effect of $dT_c/dp$ = -1.13 K/GPa. However, when pressure is released, the $T_c$ increases with the same rate as in the gas-pressure cell. Therefore, we believe that during pressure application in the DAC a degradation effect of the sample occurs which might originate in slight uniaxial stresses. Such undesirable side effects are not avoidable when liquid or solid pressure transmitting media are used at low temperature in a DAC. However, the degradation effect seems to occur only during pressure application, whereas with decreasing pressure, the same pressure effect as in the gas-pressure experiments is found. The degradation is irreversible even after a complete release of the pressure.

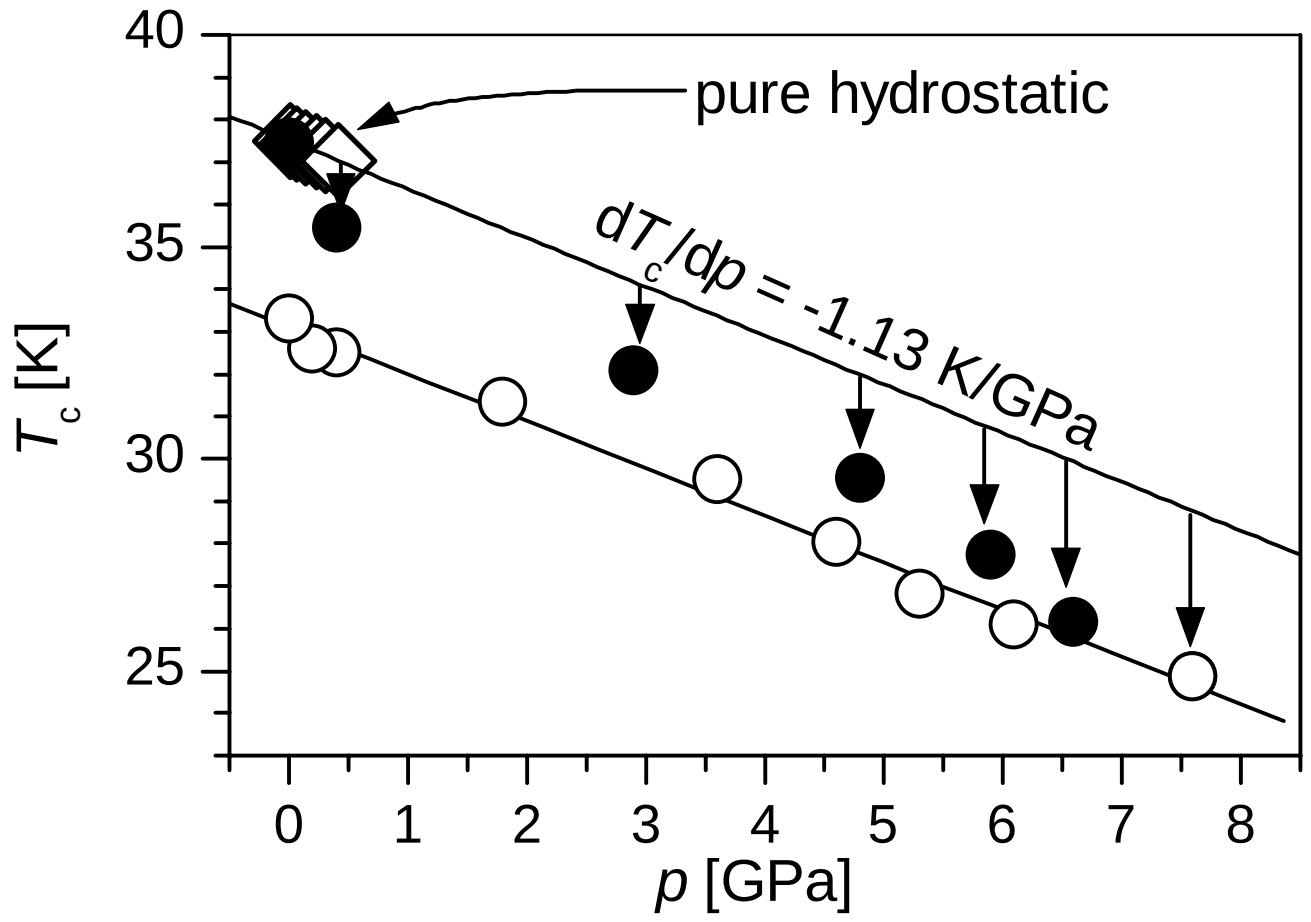

**FIGURE 6:** $T_c$ of $MgB_2$ powder samples as a function of pure hydrostatic and quasi-hydrostatic pressure, measured in the gas-pressure cell (◊) and diamond-anvil cell (during application (•) and release (o) of pressure), respectively.

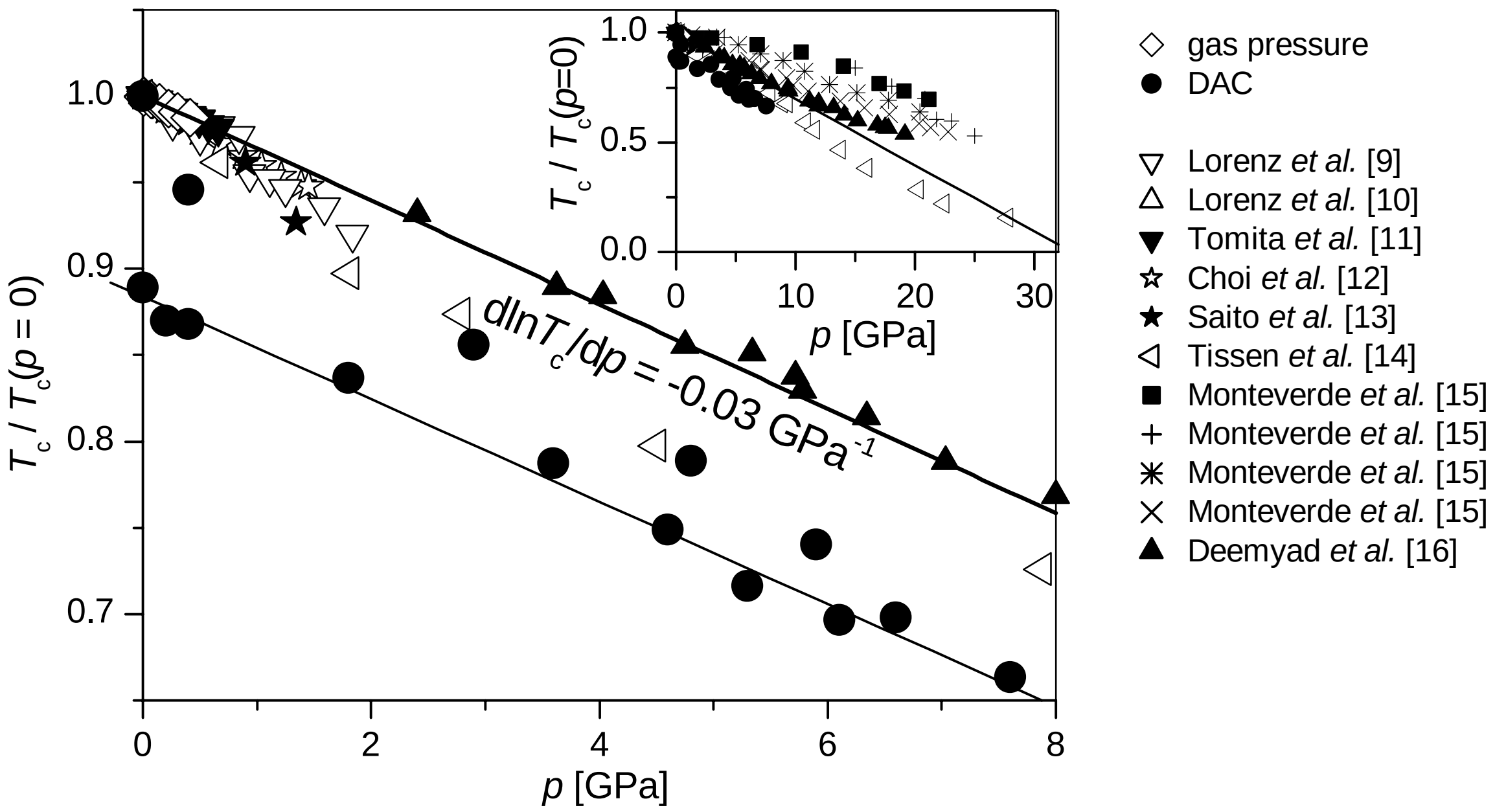

**FIGURE 7**: $T_c$ versus pressure. Comparison of our results to literature data.

When we compare our $T_c(p)$ results to the measurements reported in the literature (FIG. 7) [9-17], we find that all pressure effects determined with helium as pressure-transmitting medium show $T_c$-decreases similar to the value we determined in our gas-pressure cell ($dT_c/dp$ = –1.13 K/GPa). In the measurements with solid or liquid pressure-transmitting media stronger $T_c$-decreases were found (with the exception of the results of Monteverde *et al.* [15] who determined $T_c$ as the onset of the resistive transition, which might not directly be comparable to transition temperatures determined by ac-susceptibility measurements and/or by the 50% criterion). Therefore, we believe that the intrinsic pressure effect under purely hydrostatic pressure conditions is approximately $dT_c/dp$ = -1.13 K/GPa and that the anisotropic structure of $MgB_2$ seems to be very sensitive to uniaxial pressure components.

## CONCLUSIONS

We measured $T_c$ and lattice parameters of $MgB_2$ powder as a function of pressure. $MgB_2$ was found to be a very hard material with a bulk modulus of $B_0$ = 196 GPa. The *c*-axis is much softer than the *a*-axis, which means that hydrostatic pressure compresses the out-of-plane Mg-B bonds more effectively than the bonds within the boron layers. This corresponds with a strong anisotropic thermal expansion being twice as high for the *c*-axis compared to the *a*-axis [5].

In the $T_c$-measurements under purely hydrostatic pressure conditions up to 0.42 GPa we observed a linear decrease of $T_c$ with $dT_c/dp$ = -1.13 K/GPa. The measurements in the diamond anvil cell revealed a degradation of the sample during pressure application with a strong $T_c$-decrease of $dT_c/dp$ = -1.6 K/GPa. This degradation effect is explained by a large sensitivity of the anisotropic, layered structure to small uniaxial pressure components within the sample chamber, which cannot be avoided at high pressures and low temperatures using a diamond-anvil cell with liquid or solid pressure transmitting media. Such a degradation effect might be an explanation for the spread of $dT_c/dp$ values reported in the literature.